\documentclass[aps,prd,preprint,groupedaddress,showpacs,11pt]{revtex4-1}
\usepackage{amsmath,amstext,amsbsy,amssymb}
\usepackage{bm}

\newcommand{\Tr}{{\rm Tr}}

\newcommand{\cme}{{\scriptscriptstyle{ CME}}}
\newcommand{\cve}{{\scriptscriptstyle{ CVE}}}
\newcommand{\ssA}{{\scriptscriptstyle{ A}}}
\newcommand{\ssV}{{\scriptscriptstyle{ V}}}
\newcommand{\ssD}{{\scriptscriptstyle{ D}}}

\newcommand{\ssR}{{\scriptscriptstyle{ R}}}
\newcommand{\ssH}{{\scriptscriptstyle{ H}}}
\newcommand{\ssS}{{\scriptscriptstyle{ S}}}
\newcommand{\ssF}{{\scriptscriptstyle{ F}}}
\newcommand{\ssE}{{\scriptscriptstyle{ E}}}

\newcommand{\ssL}{{\scriptscriptstyle{ L}}}

\newcommand{\ssM}{{\scriptscriptstyle{ M}}}
\newcommand{\ssC}{{\scriptscriptstyle{ C}}}
\newcommand{\ssP}{{\scriptscriptstyle{ P}}}

\newcommand{\ssh}{{\scriptscriptstyle{1/2}}}

\newcommand{\sif}{{\scriptscriptstyle{ 0}}}

\long\def\symbolfootnote[#1]#2{\begingroup%
\def\thefootnote{\fnsymbol{footnote}}\footnote[#1]{#2}\endgroup}

\begin{document}
\title{ Semiclassical dynamics of Dirac and Weyl particles in rotating coordinates}

\author{\"{O}mer F. Dayi}
\email{dayi@itu.edu.tr }
\author{Eda Kilin\c{c}arslan}
\email{edakilincarslann@gmail.com}
\author{Elif Yunt}
\email{yunt@itu.edu.tr }
\affiliation{%
	Physics Engineering Department, Faculty of Science and
	Letters, Istanbul Technical University,
	TR-34469, Maslak--Istanbul, Turkey}


\date{\today}

\begin{abstract}
The semiclassical kinetic theory of Dirac  particles in the presence of external electromagnetic fields and global rotation is established. To provide the Hamiltonian formulation of Dirac particles a  symplectic two-form which is a matrix in spin indices is proposed.   The  particle number and current densities for the Dirac particles are acquired in the helicity basis. Following a similar procedure,  semiclassical kinetic theory of the  Weyl particles is accomplished. It is shown that the phase-space dynamics of the Weyl and Dirac particles  is directly linked. The anomalous chiral effects due to the external electromagnetic fields and angular velocity of the frame are calculated.
\end{abstract}

\maketitle

\section{Introduction}
\label{int}
 Quantum mechanical calculations revealed that one cannot conserve both the vector and axial currents which are originated from spin-1/2 particles, even in the vanishing mass limit due to axial anomaly. In heavy ion collisions this anomaly can yield observable  effects like the chiral magnetic effect \cite{kmw,fkw,kz} and the chiral separation effect \cite{mz,jkr}. 
There are also  similar anomalous effects due to global rotation of Fermi liquid which are known as
 the chiral vortical effect \cite{ss} and local (spin) polarization effect \cite{lw,bpr,glpww}. 
Experimental evidences of these anomalous chiral effects in relativistic heavy ion collisions were recently discussed  in \cite{khetal-arx}, where a complete list of references can be found. 

The massless Dirac equation is invariant under chiral transformations. However,  when the external  electromagnetic fields are coupled to the massless spin-1/2 particles, chiral invariance is lost due to quantum mechanical effects. 
On the other hand when the Lorentz force on a particle of charge $q,$ and mass $m,$ is expressed in a rotating coordinate frame, the Coriolis term disappears if the angular velocity of the rotation $\bm \Omega$ and the magnetic field $\bm B$ 
are related as $2mc \bm \Omega =-q\bm B.$ Hence, global rotations and magnetic fields generate similar effects
as far as the centrifugal force is ignored. This resemblance  can be extended to the massless fermion i.e. the Weyl particle.
These one particle properties can be generalized to many particles within the kinetic theory. It is possible to incorporate chiral anomaly into the classical kinetic theory of chiral particles  by adding the first order quantum corrections  \cite{soy,sy}. This semiclassical approach  yields an intuitive understanding of the chiral magnetic and chiral vortical effects. Semiclassical chiral kinetic theory  was also studied within the Hamiltonian formalism using symplectic two-forms \cite{ds}  by associating some classical variables to spin. 

Noninteracting, massive spin-1/2 particles obey the Dirac equation which yields two  Weyl equations in the massless case.  These are relativistic systems where the energy can be positive or negative. Although
a complete quantum mechanical wave packet should be formed by  positive as well as negative energy  solutions,  to get an intuitive picture of some quantum phenomena, it is possible to consider wave packets which
are composed of positive energy solutions only \cite{sundaramniu,Culcer,chni,ccn}. This is the semiclassical  approach which we deal with. This formalism unlike the others include the Berry connection obtained from the free particle solutions. Within the semiclassical wave packet formalism the Berry gauge fields cease to be  pure gauge fields, so that they yield a nonvanishing curvature. A combination of 
the differential form method of \cite{ds} with the semiclassical wave packet formalism was presented in \cite{de}. In this approach the spin degrees of freedom are kept  explicit, so that  the symplectic two-form  which is a matrix in spin indices is introduced. This method was proved to be efficient in deriving  the chiral magnetic effect and the chiral anomaly. Within this formalism  some aspects of the semiclassical kinetic theory of the Dirac particles in the presence of electromagnetic fields was analyzed in \cite{om-eda} (for another approach see \cite{mwp}). Moreover, it was  demonstrated that the differential form formalism  in terms of matrices in spin indices  is crucial  to express the spin Hall conductivity in terms of the topological spin Chern number of the systems obeying Dirac-like equations \cite{om-elif}. 

We would like to study the semiclassical kinetic theory of the  Dirac and Weyl particles in the presence of the external electromagnetic fields in a uniformly rotating coordinate frame by keeping the usually ignored centrifugal force terms. There are several reasons of dealing with rotating coordinates. First of all, as it has been emphasized in \cite{HehlNi} where the Dirac particle was studied in a rotating coordinate frame, the laboratories on Earth are affected by the rotation of Earth. By observing a fluid element in a frame rotating with respect to laboratory frame, one can study vortical effects \cite{sy}.
From the high energy point of view it is important to analyze chiral particle currents which can produce anomalous chiral transport effects. However, some aspects of the  semiclassical kinetic  theory   of chiral particles like the dispersion relation \cite{chenetal}, can be established systematically  by studying the massive case \cite{mtr}. 
Because of considering both the external electromagnetic fields and global rotation we can keep track of the similarities and differences between the effects generated by them.  Moreover, we can handle systematically how they affect each other and if they furnish some joint phenomena.

Spin-dependent interactions of Dirac particles may yield chiral imbalance of the chiral particles. Our method provides a direct relation between the spin-dependent interactions of Dirac fermions with the chiral particles. It constitutes the first step in the  systematic study  of  interactions between the right- and left-handed fermions.
On the other hand the semiclassical kinetic theory  the  Dirac particle in rotating coordinate frames may find applications in condensed matter systems where mechanical rotations of Dirac particles  can generate spin currents  \cite{mism,mism2}.  
Obviously   rotation which we consider is not an intrinsic property of the system, so that the rotating coordinate system which we consider is not necessarily Lorentz invariant.  

The Dirac equation in the presence of external electromagnetic fields and mechanical rotation was also studied in \cite{mism,mism2}. By considering the
low-energy limit of  the Dirac Hamiltonian they derived the related Pauli-Schr\"{o}dinger equation which has only positive energy solutions given by  the unit spinors $(1\ 0)^T,\ (0\ 1)^T.$
In the absence of rotation, the semiclassical transport of fermions based on the Pauli-Schr\"{o}dinger equation has been investigated in \cite{chni}. They showed  that  the formalism based on positive energy solutions of the Dirac equation  and the one derived from the  Pauli-Schr\"{o}dinger equation are different. In \cite{mism2} wave packet formalism was not discussed. Nevertheless, they proposed semiclassical equations of motion in terms of the force they had calculated. To have an idea about the differences between  their proposal and our approach, in  Appendix we present the force arising in our approach.

Wigner functions are quantum mechanical analogs of classical distribution functions.  The Wigner function constructed by Dirac spinor fields satisfies a quantum  kinetic equation \cite{qBe}. In \cite{glpww} a solution of the quantum kinetic equation was obtained  for weak external fields. Within this approach a Lorentz covariant Boltzmann (kinetic) equation was presented in \cite{cpww} for chiral particles. By integrating it  over the energy (zeroth component of the 4-momentum) they obtained a solution for the first time derivatives  of thre-dimensional phase-space variables. Although, first time derivatives of spatial coordinates which they propose are similar to the ones which we acquired, solutions for the measure and first time derivatives of momentum variables differ considerably as we will discuss in Sec. \ref{trawe}. 

The free Dirac Hamiltonian does not commute with spin operator but it commutes with the helicity operator.
Thus helicity is a conserved quantity under the time evolution generated by the free Dirac Hamiltonian.
Hence it is natural to work in  helicity basis to obtain the particle number and current densities. On the other hand chirality is equal to helicity for the massless Dirac equation. Thus, in the helicity basis the effects of imbalance between the right- and left-handed particles can be studied explicitly.

The semiclassical kinetic equations of Dirac and Weyl particles which we established by including the rotation of coordinates  beginning from the nonrelativistic particles are novel. In fact, symplectic two forms which we introduce to construct the semiclassical equations by making use of differential form approach are new.  We show that our results for the Weyl particles are consistent with the hydrodynamic approach when we deal with  Boltzmann equation without collisions. However the main power of our kinetic equations  will be clear when we deal with collisions which can directly be introduced within our approach for instance by adopting the relaxation time approximation. Only in the presence of collisions particle currents will acquire some new contributions. These are currently under consideration.

The paper is organized as follows. In the next section we discuss the main ingredients of our semiclassical approach. We present how the Berry curvature arises naturally in the semiclassical wave packet formalism. In Sec. \ref{sfrc} the symplectic matrix two-form which is suitable to establish semiclassical formulation of the Dirac particles coupled to external  electromagnetic fields in rotating coordinates  is presented. Sec. \ref{Semiclassical Dirac Hamiltonian in rotating coordinates} is devoted to obtain the related Hamiltonian including  terms at the order of Planck constant. In Sec. \ref{trans} we derived continuity equation for the Dirac particles. Semiclassical kinetic theory of the Weyl particles coupled to electromagnetic fields in rotating coordinates is developed in Secs.  \ref{hamwe} and \ref{trawe}. 
In Sec. \ref{ace} we present anomalous chiral effects  arising in our formalism. They are in accord with the chiral effects obtained within other formalisms. The results acquired and their possible applications are discussed in the last section.

\section{ Semiclassical Approximation and the Berry Curvature }
\label{sember}

Our semiclassical approach  is based on the two linearly independent positive energy solutions  
  of the Dirac equation furnished by the Dirac Hamiltonian 
\begin{equation}
H_\ssD^{\scriptscriptstyle{(4)}} (\bm{p})=\beta m + \bm{\alpha} \cdot \bm{p}.
\label{hamiltonian}
\end{equation}
We set the speed of light $c=1$ and choose the following representation of
 $\beta,\ \alpha_i;\ i=1,2,3,$  matrices,
$$
\bm \alpha=
\begin{pmatrix}
0 & \bm \sigma \\
\bm \sigma & 0
\end{pmatrix},\qquad
\beta=
\begin{pmatrix}
1 & 0\\
0 & -1
\end{pmatrix} ,
$$
where
$\bm \sigma$ are  the Pauli spin matrices.  
The semiclassical wave packet is defined by means of  the positive energy solutions $u^\alpha ( \bm{p});\ \alpha=1,2,$ as 
$$
\psi_{\bm x} (\bm{p}_c) = \sum_\alpha \xi_\alpha  u^\alpha  (\bm{p}_c) e^{-i\bm{p}_c \cdot \bm{x} /\hbar}.
$$ 
The coefficients  $\xi_\alpha , $ are chosen to be constant.
$\bm{x}_c,$ and  $\bm{p}_c,$ denote the phase-space  coordinates of wave packet center coinciding with the center of mass.  We define the one-form $\eta_0$ through 
$$
 \int [dx] \delta(\bm{x}_c - \bm{x}) \Psi^\dagger_{\bm{x}}\left( -i \hbar d -H^{\scriptscriptstyle{(4)}}_\ssD dt \right)\Psi_{\bm{x}}=\sum_{\alpha\beta}\xi^*_{\alpha} \eta^{\alpha\beta}_0 \xi_{\beta}.
$$ 
$\eta_0,$ which is a matrix in the ``spin indices"  $\alpha, \beta ,$  can be written as
\begin{equation}
\eta^{\alpha\beta}_0= - \delta^{\alpha\beta}\bm{x}_c\cdot d\bm{p}_c  - \bm A^{\alpha\beta}\cdot d{\bm p}_c -H_\ssD^{\alpha\beta}dt  .
\label{et1}
\end{equation}
Here $H_\ssD^{\alpha\beta}$ is the projection of the Dirac Hamiltonian (\ref{hamiltonian})  on the positive energy solutions. Moreover, we introduced the matrix valued Berry gauge field 
\begin{equation}
\label{bgd}
\bm A^{\alpha\beta}= -i \hbar u^{\dagger(\alpha)}(\bm p_c)\frac{\partial }{\partial {\bm p_c}} u^{(\beta)}(\bm p_c). 
\end{equation}
By relabeling $(\bm{x}_c,\bm{p}_c)\rightarrow (\bm{x},\bm{p})$ and adding an exact differential term, the one-form (\ref{et1}) can be rewritten as
$$
\eta_0=\bm p \cdot d\bm x - \bm A \cdot d\bm p -H_\ssD dt.
$$
Unless necessary the spin indices and the related unit matrix are suppressed. 
Before proceeding with the Hamiltonian formulation let there be external vector fields, like electromagnetic fields, which are provided  by  the gauge field $\bm a (\bm x, \bm p , t )$ and the scalar field $\phi ( \bm x,  \bm p ,t ).$  
The presence of external fields will also alter the starting Hamiltonian. Because of dealing with
the semiclassical formulation up to first order in the Planck constant, the Hamiltonian would be decomposed as $H^{\scriptscriptstyle{(4)}}=H^{\scriptscriptstyle{(4)}}_0 +\hbar H^{\scriptscriptstyle{(4)}}_1.$

Let us 
consider the first-order Hamiltonian formalism designated by the one-form
\begin{equation}
\eta=\bm p \cdot d\bm x - \bm A  (\bm p  ) \cdot d\bm p  +\bm a  (\bm x, \bm p , t ) \cdot d\bm  x +\phi  (\bm x, \bm p , t )dt -H  (\bm x, \bm p , t ) dt,
\label{eta}
\end{equation} 
where $H^{\alpha \beta}\equiv H, $ denotes the projection of $H^{\scriptscriptstyle{(4)}}$  on the positive energy solutions of the free Dirac equation.
We need to introduce  the related symplectic two-form to establish the  Hamiltonian formulation. The Berry gauge field $ \bm A  (\bm p  ) ,$ can be non-Abelian. Moreover,  we also consider  $\bm a (\bm x, \bm p , t ),\ \phi ( \bm x,  \bm p ,t ),$  which can be noncommuting with $ \bm A  (\bm p  ) .$   Therefore we define  the symplectic two-form matrix by
$$
\tilde{\omega}_t = d\eta \equiv dt \frac{\partial \eta }{\partial t} + d \bm x \cdot \frac{\partial \eta }{\partial \bm x }
 +  d\bm p \cdot \bm D \eta
$$
where we introduced the covariant derivative
$$
\bm D \equiv \frac{\partial }{\partial \bm p}+\frac{i}{\hbar}[\bm A,\ ].
$$
By employing the one-form  (\ref{eta}), we acquire
\begin{eqnarray}
\tilde{\omega}_t &=& {dp}_i \wedge{dx}_i +D_i a_j\ {dp}_i \wedge{dx}_j- G +F + \left(\frac{\partial \phi}{\partial x_i} - \frac{\partial a_i}{\partial t}\right)\  {dx}_i \wedge dt -\frac{\partial H}{\partial x_i}\  {dx}_i \wedge dt \nonumber\\
&+& D_i \phi \ {dp}_i \wedge dt  -D_i H\ {dp}_i \wedge dt.
\end{eqnarray}
As usual the repeated indices are summed over.
$F$ is the  curvature two-form of the  gauge field $\bm a,$ 
$$
F=\frac{1}{2}\left( \frac{\partial a_j}{\partial x_i}-\frac{\partial a_i}{\partial x_j}\right){dx}_i \wedge {dx}_j,
$$
and the Berry curvature two-form  $G=\frac{1}{2} {G_{ij}}{dp}_i \wedge {dp}_j$  is defined through the covariant derivative,  
\begin{equation}
G_{ij} =-i\hbar[D_i.D_j]=\left( \frac{\partial A_j}{\partial p_i}- \frac{\partial A_i}{\partial p_j}+\frac{i}{\hbar}[A_i,A_j]\right)= {\epsilon}_{ijk}G_k
\label{eq:G}.
\end{equation}
Let us calculate the Berry gauge field (\ref{bgd}), and the Berry curvature (\ref{eq:G}).
The positive energy solutions of the Dirac equation can be written as
\begin{equation}
\label{ubas}
u^\alpha (\bm{p}) = U_0(\bm{p}) {u_0}^\alpha,
\end{equation}
where 
${u_0}^{1}=(1\ 0\ 0\	0)^{T}, \ \ {u_0}^{2}=(0\ 1\ 0\	0)^{T},$  are the rest frame solutions and  $U_0(\bm{p})$ is the Foldy-Wouthuysen transformation,
\begin{equation}
\label{FWt}
U_0(\bm{p}) =\frac{\beta H_\ssD^{\scriptscriptstyle{(4)}} (\bm p)+ E}{\sqrt{2E (E+m)}}.
\end{equation}
$E=\sqrt{p^2+m^2}$ is the free  relativistic energy.
Now, the Berry gauge field (\ref{bgd}) is expressed as
$$
\bm A = -i\hbar  I_+  U_0(\bm{p}) \frac{\partial U_0^\dagger(\bm p)}{\partial \bm p}  I_+ .
$$
$I_+$ projects onto the positive energy subspace. Hence,  we acquire
$$
\bm{A}=\hbar \frac{\bm{\sigma} \times \bm{p}}{2E(E+m)},
$$
and by plugging it into (\ref{eq:G}) one establishes
\begin{equation}
\bm G= \frac{\hbar m}{2E^3}\left( \bm{\sigma}+\frac{\bm{p}(\bm{\sigma}\cdot\bm{p})}{m(m+E)}\right)
\label{berrycurvature}.
\end{equation}
It furnishes the Berry curvature via $ G_{ij} ={\epsilon}_{ijk}G_k.$

\section{Symplectic Two-form  Matrix in Rotating Coordinates}
\label{sfrc}

We aim to study the Dirac particle in the presence of external electromagnetic fields, in the  coordinate frame
rotating with the  constant angular velocity  $\bm \Omega .$
Nonrelativistic
global rotations, i.e.  which fulfill $|\bm \Omega \times \bm x|\ll c,$  can be associated with the vector gauge field $m (  \bm \Omega  \times \bm x),$ and with the scalar gauge field $\frac{m}{2}(\bm \Omega \times \bm x)^2$ \cite{rotgf}, for a nonrelativistic particle of mass $m.$  We can extend them to our relativistic particle  formulation by $\bm a^\Omega={\cal E}( \bm \Omega  \times \bm x ),$ and  $\phi^\Omega =\frac{\cal E}{2}(\bm \Omega \times \bm x)^2,$ where 
${\cal E} \equiv H$ is  the  dispersion relation which yields $m,$ in the nonrelativistic limit.
We then set  $\bm a= \bm a^{\ssE \ssM} + \bm a^\Omega $ and $\phi =\phi^{\ssE \ssM}+\phi^\Omega $ in the one-form (\ref{eta}), so that we deal with the one-form matrix,
\begin{equation}
\eta=\bm p \cdot d\bm x - \bm A \cdot d\bm p  +\bm a^{\ssE \ssM}  \cdot d\bm  x + {\cal E}( \bm \Omega  \times \bm x )\cdot d\bm  x +\phi^{\ssE \ssM} dt +\frac{\cal E}{2}(\bm \Omega \times \bm x)^2 dt -H dt,
\label{etarot}
\end{equation} 
Equation (\ref{etarot}) furnishes the symplectic two-form needed for the Hamiltonian formalism
in rotating coordinates as
\begin{eqnarray}
\tilde{\omega}_t & = & {dp}_i \wedge {dx}_i + \frac{1}{2} \epsilon_{ijk} (q B_k + 2 {\cal E} \Omega_k)\ {dx}_i \wedge {dx}_j -\frac{1}{2} \epsilon_{ijk} G_{k}\ {dp}_i \wedge {dp}_j \nonumber \\ 
&&+ \epsilon_{ijk} x_j\Omega_k    \nu_m {dx}_i \wedge {dp}_m - \nu_i\ {dp}_i \wedge dt + \frac{1}{2}  \nu_i (\bm \Omega \times \bm x)^2 {dp}_i \wedge dt \nonumber\\
&&+[q\bm E+ (\bm\Omega\times \bm x)\times (q\bm B +{\cal E} \bm\Omega)]_i\ {dx}_i\wedge dt \label{wtf}.
\end{eqnarray}
The  electromagnetic vector and scalar  potentials, $\bm a^{\ssE \ssM}, \phi^{\ssE \ssM},$ are chosen appropriately to get the terms depending on the external electric and magnetic fields, $\bm E,  \bm B,$ as in (\ref{wtf}).  Moreover,  we introduced
\begin{equation}
\label{vufM}
\bm \nu=\frac{\partial H }{\partial \bm p}+\frac{i}{\hbar}[\bm A, H ] \equiv \bm D H .
\end{equation} 
The Hamiltonian   $H\equiv {\cal E}$ will be presented in the next section.  In (\ref{etarot}) we  generalized the  nonrelativistic fields, $m (  \bm \Omega  \times \bm x),\ \frac{m}{2}(\bm \Omega \times \bm x)^2,$ which depend explicitly on $\bm x,$  by substituting $m,$  with ${\cal E},$ so that  the fourth and sixth terms appear in the generalized two-form (\ref{wtf}). These terms are essential to construct the semiclassical kinetic equation correctly.

The equations of motion can be derived by imposing the condition
\begin{eqnarray}
i_{\tilde{v}} \tilde{\omega}_t = 0,
\label{i_v}
\end{eqnarray}
where $i_{\tilde{v}}$ denotes the interior product of the vector field
\begin{equation}
\label{vf}
\tilde v= \frac{\partial}{\partial t}+\dot{\tilde {\bm x}}\frac{\partial}{\partial \bm{x}}+\dot{\tilde {\bm p}}\frac{\partial}{\partial \bm{p}}.
\end{equation}
$(\dot{\tilde {\bm x}},\dot{\tilde {\bm p}})$ are the matrix-valued time evolutions of the phase-space variables $(\bm x, \bm p).$
The equations of motion are deduced by making use of   (\ref{wtf}) in (\ref{i_v}):
\begin{eqnarray}
\dot{\tilde {\bm x}} + (\bm \Omega \times \bm x) \cdot  \dot{\tilde {\bm x}}\ \bm \nu &=& \bm \nu (1 -\frac{1}{2} (\bm \Omega \times \bm x)^2) +  \dot{\tilde {\bm p}} \times \bm G  ,\label{xeq} \\
\dot{\tilde {\bm p}} + (  \dot{\tilde {\bm p}} \cdot \bm \nu) \bm \Omega \times \bm x &=& q\bm E + (\bm\Omega\times \bm x)\times (q\bm B +{\cal E} \bm\Omega) \ +  \dot{\tilde {\bm x}} \times  (q\bm B +2 {\cal E} \bm \Omega ) .\label{peq}
\end{eqnarray}
The second and third terms in the right-hand side of (\ref{peq}), respectively, encompass the centrifugal  and  Coriolis forces correctly. The left-hand sides of (\ref{xeq}) and (\ref{peq}) resemble the Lorentz transformations of  velocity 
$\dot{\tilde {\bm x}},$ and force $\dot{\tilde {\bm p}},$ to a reference frame moving with the velocity
$\bm v=\bm \Omega \times \bm x.$
To derive the explicit expression of $\bm \nu, $ we need to
be acquainted  with the underlying Hamiltonian.  

\section{Semiclassical Dirac Hamiltonian in Rotating Coordinates}
\label{Semiclassical Dirac Hamiltonian in rotating coordinates}

To accomplish the semiclassical energy, we need to  clarify  the starting Hamiltonian in the classical phase-space variables $(\bm x , \bm p).$ The simplest choice is to deal with the free Dirac Hamiltonian. However, we are interested in the semiclassical approximation where the terms at the first order in $\hbar$  have been retained. When the Dirac particle is subject to the external magnetic field $\bm B,$ in \cite{Bliokh} Bliokh suggested  to add   the  $(-\frac{\hbar q}{2E}\bm{\Sigma}\cdot \bm{B}),$ term   to the Hamiltonian (\ref{hamiltonian}), where 
$$\bm{\Sigma}=\begin{pmatrix} \bm{\sigma}&0\\0&\bm{\sigma}\end{pmatrix},$$
is the spin matrix.
Observe that this additional term can be acquired from the magnetic moment-magnetic field  interaction term of an electron by the substitution of mass $m,$ with the relativistic free energy $E.$ By making use of the analogy between magnetic field and angular velocity one can write the following Hamiltonian 
\begin{equation}
\label{reham}
H^{\scriptscriptstyle{(4)}}= \beta m +\bm{\alpha}\cdot\bm{ p }  -\frac{g\hbar}{2}\bm{\Sigma}\cdot \bm{\Omega}-\frac{\hbar q}{2E}\bm{\Sigma}\cdot \bm{B}.
\end{equation}
We introduced the constant $g,$ which  should be set as $g=2,$ to keep the analogy between  magnetic field and angular velocity which shows up in the Coriolis term in (\ref{peq}). However, the Dirac Hamiltonian in a  rotating coordinate frame  was established in \cite{HehlNi}, where the  spin-angular velocity coupling term appears with $g=1.$ For the sake of generality, we  retain $g.$ 

The Hamiltonian  (\ref{reham}) can be decomposed into two parts: $H^{\scriptscriptstyle{(4)}}=H^{\scriptscriptstyle{(4)}}_\ssD +\hbar H^{{(4)}}_1$.  
The semiclassical Hamiltonian is defined to be the  projection of the Hamiltonian (\ref{reham})  onto the positive energy solutions. Instead of doing this calculation straightforwardly  we would like to first  block diagonalize   $H^{\scriptscriptstyle{(4)}}$ and then project it onto the positive energy subspace. 
In order to block diagonalize  (\ref{reham}) up to   $\hbar$ order, let us introduce the transformation 
$U_0+\hbar U_1,$  yielding
\begin{equation}
\label{u0u1}
(U_0+\hbar U_1)(H^{\scriptscriptstyle{(4)}}_\ssD +\hbar H^{\scriptscriptstyle{(4)}}_1)(U_0+\hbar U_1)^{\dagger}=U_0H^{\scriptscriptstyle{(4)}}_\ssD U^{\dagger}_0 +\hbar(U_0H^{\scriptscriptstyle{(4)}}_\ssD U_1^{\dagger}+U_0H^{\scriptscriptstyle{(4)}}_1U_0^{\dagger}+U_1H^{\scriptscriptstyle{(4)}}_\ssD U_0^{\dagger}) +O(\hbar^2) .
\end{equation}
$U_0(\bm p)$ denotes the Foldy-Wouthuysen transformation given in (\ref{FWt}). It diagonalizes the Dirac Hamiltonian (\ref{hamiltonian}) and acts on the spin dependent part as 
\begin{equation}
\label{ofdi}
 U_0H^{\scriptscriptstyle{(4)}}_1U_0^{\dagger}=-\frac{ m}{2E^2}(q\bm B+gE\bm \Omega)\cdot\bm{\Sigma}- \frac{i}{2E^2}\bm{ p }\cdot\left[(q\bm B+gE\bm \Omega)\times\beta\bm{\alpha}\right]-\frac{1}{2E^2(E+m)}\bm{\Sigma}\cdot\bm{ p }\ (q\bm B+gE\bm \Omega)\cdot\bm{ p }.
 \end{equation}
To get rid of the off-diagonal blocks  we choose 
 $$
 U_1=\frac{1}{2E^3\sqrt{2E(E+m)}} \begin{pmatrix} i\bm{\sigma}\cdot \left[\bm{ p }\times(q\bm B+gE\bm \Omega)\right](\bm{\sigma}\cdot\bm{ p })&-i(E+m)\bm{\sigma}\cdot\left[\bm{ p }\times(q\bm B+gE\bm \Omega)\right]\\ -i(E+m)\bm{\sigma}\cdot\left[\bm{ p }\times(q\bm B+gE\bm \Omega)\right]& -i\bm{\sigma}\cdot\left[\bm{ p }\times(q\bm B+gE\bm \Omega)\right](\bm{\sigma}\cdot\bm{ p })\end{pmatrix}.$$
One can indeed show that 
 the off-diagonal blocks in (\ref{ofdi}) are canceled by $(U_1H^{\scriptscriptstyle{(4)}}_\ssD U_0^{\dagger}+U_0H^{\scriptscriptstyle{(4)}}_\ssD U_1^{\dagger}).$ 
Therefore (\ref{u0u1}) provides the block-diagonal Hamiltonian in the rotating frame as follows
\begin{eqnarray}
\label{bdhh}
H^{\scriptscriptstyle{(4)}}=E\beta - \frac{\hbar m}{2 E^2} \bm{\Sigma} \cdot (q\bm B+gE\bm \Omega) - \frac{\hbar}{ 2E^2(E+m)} (\bm{\Sigma} \cdot \bm{ p }) (q\bm B+gE\bm \Omega) \cdot \bm{ p }.
\end{eqnarray}
For  $\bm \Omega =0,$ this block-diagonal Hamiltonian 
was  established   in  \cite{Bliokh}, which had been found  also  in \cite{GBM} within a   systematic but  somewhat complicated approach.

Projection of (\ref{bdhh})  onto the positive energy subspace  furnishes  the desired Hamiltonian:
\begin{equation}
\label{Hambo}
H=E  - \frac{\hbar m}{2 E^2} \ \bm{\sigma} \cdot (q\bm{B} + gE\bm{\Omega}) - \frac{\hbar}{2 E^2(E+m)} (\bm{\sigma} \cdot \bm{p}) \ \bm{p} \cdot (q\bm{B} + gE\bm{\Omega}).
\end{equation} 
Observe  that it is indeed the projection of the initial Hamiltonian, (\ref{reham}), onto the positive energy solutions of the Dirac equation. Equation (\ref{Hambo}) can be expressed as
\begin{equation}
\label{hwbc}
H=E [1-\bm G \cdot (q\bm{B} + gE\bm{\Omega}) ]
\end{equation}
by employing  the Berry curvature (\ref{berrycurvature}).

\section{Semiclassical Transport of Dirac Particles}
\label{trans}

To analyze the  particle number conservation law we start with   the volume form
\begin{eqnarray}
\label{vftw}
\tilde{\Omega} &=& \frac{1}{3!} \tilde{\omega}_t \wedge \tilde{\omega}_t \wedge \tilde{\omega}_t \wedge dt \nonumber\\
&=& 	\frac{1}{3!} \tilde{\omega} \wedge \tilde{\omega} \wedge \tilde{\omega} \wedge dt .
\end{eqnarray}
$\tilde{\omega}\equiv \tilde{\omega}_t|_{dt=0}$ is the matrix valued symplectic two-form  in the ordinary phase-space whose  coordinates are $(\bm x,\bm p).$ The volume form
(\ref{vftw}) can be expressed as 
\begin{equation}
\label{wfpf}
\tilde{\Omega}= \tilde{\omega}_\ssh \ dV \wedge dt,
\end{equation}
where $\tilde{\omega}_\ssh $  is the Pfaffian of the $(6\times 6)$ matrix,
\begin{equation}
\label{syma}
\begin{pmatrix}
\epsilon_{ijk} (q B_k + 2E \Omega_k) & -\delta_{ij}+\nu_j(\bm x \times\bm \Omega)_i \\
\delta_{ij}-\nu_i(\bm x \times\bm \Omega)_j &\ -\epsilon_{ijk} G_{k}
\end{pmatrix}. 
\end{equation}
We will accomplish  the explicit form of the Pfaffian $\tilde{\omega}_\ssh$ in the sequel.
To attain the Liouville equation, we need to calculate the Lie derivative of  the volume form which can be carried out in two different ways. One of them is to utilize the definition of the volume form in terms of the Pfaffian, (\ref{wfpf}):
\begin{eqnarray}
{\cal{L}}_{\tilde v} \tilde{\Omega} &=& (i_{\tilde v}  d + d i_{\tilde v} ) (\tilde{\omega}_\ssh dV \wedge dt) \nonumber\\
&=& \left( \frac{\partial \tilde{\omega}_\ssh}{\partial t} + \frac{\partial}{\partial \bm x}\cdot  ( \dot{\tilde{ \bm x}}\tilde{\omega}_\ssh) + \bm D\cdot (\tilde{\omega}_\ssh \dot{\tilde {\bm p}})\right) dV \wedge dt.
\label{lievolume2}
\end{eqnarray}
The other way is to employ  the  definition of  volume form  (\ref{vftw}) and directly compute its Lie derivative:
\begin{eqnarray}
{\cal{L}}_{\tilde v} \tilde{\Omega} &=& (i_{\tilde v}  d + d i_{\tilde v} )(\frac{1}{3!} \tilde{\omega}_t^3 \wedge dt)\nonumber\\
&=& \frac{1}{3!} d {\tilde{\omega}_t}^3 .
\label{liouville}
\end{eqnarray}
Explicit calculation of ${\tilde{\omega}_t}^3$ and the comparison of (\ref{liouville})  with  (\ref{lievolume2}), provide us   the explicit form of  Pfaffian and $ \dot{\tilde{ \bm x}}\tilde{\omega}_\ssh ,$ $\tilde{\omega}_\ssh \dot{\tilde {\bm p}},$ which are the solutions of the equations of motion  (\ref{xeq})-(\ref{peq}), in terms of the phase-space variables  $(\bm x,\bm p),$ as
\begin{eqnarray}
\tilde{\omega}_{\ssh} &=&1+  \ \bm{G} \cdot (q\bm{B} + 2 {\cal E} \bm{\Omega} ) - \bm \nu \cdot (\bm x \times \bm \Omega) - (  \bm \nu  \cdot \bm G)(q \bm B \cdot (\bm x \times \bm \Omega)),
\label{Pfaf} \\
\dot{\tilde{ \bm x}}\tilde{\omega}_\ssh&=&{\bm \nu} (1 -\frac{1}{2} (\bm \Omega \times \bm x)^2) +  \ \bm e \times \bm G \nonumber\\
&&+  ({\bm \nu} \cdot \bm G) (q \bm B+ 2{\cal E} \bm \Omega)(1 -\frac{1}{2} (\bm \Omega \times \bm x)^2)+ ( \bm \nu \cdot \bm G) [ (\bm x \times \bm \Omega) \times \bm e ]    ,\label{msxd}\\
\tilde{\omega}_\ssh \dot{\tilde {\bm p}}&=& \bm e + {\bm \nu} \times (q\bm{B} + 2 {\cal E} \bm{\Omega}) (1 -\frac{1}{2} (\bm \Omega \times \bm x)^2) \nonumber\\
&& + \bm G (\bm e \cdot (q \bm{B} + 2{\cal E} \bm{\Omega})) - [ (\bm x \times \bm \Omega) \times \bm e ] \times {\bm \nu} .  \label{mspd}
\end{eqnarray}
$\bm e$ denotes the effective electric field:
\begin{equation}
\label{efel}
\bm e =  q\bm E + (\bm\Omega\times \bm x)\times (q\bm B +{\cal E}\bm\Omega). 
\end{equation}
The second term in  (\ref{mspd}), reflects the fact that ${2\cal E}\bm \Omega$ behaves as an effective magnetic field in the classical limit.  By
 plugging  the semiclassical Hamiltonian (\ref{Hambo}), into the definition (\ref{vufM}),  one calculates
 $\bm \nu$ as 
\begin{equation}
\bm \nu =\frac{\bm p}{E} \left[ 1
+2\bm G \cdot \left(  q \bm B+g\frac{E}{2}\bm{\Omega}\right) \right]
- \frac{ \hbar }{2E^3}   \ (q\bm B +gE\bm \Omega) \bm{\sigma} \cdot \bm{p}  . \label{numas}
\end{equation}
 
Direct computation of the Lie derivative, (\ref{liouville}), leads to 
\begin{eqnarray}
&(\frac{1}{2}d\tilde{\omega}_t \wedge \tilde{\omega}_t^2 )_{\ssV\ssM} &= 
 (1 -\frac{1}{2} (\bm \Omega \times \bm x)^2)(q\bm B+2E\bm \Omega) \cdot
(\bm{D}\times \bm{\nu})  \nonumber \\
&& =(1 -\frac{1}{2} (\bm \Omega \times \bm x)^2) (q\bm B+2E\bm \Omega)\cdot \left( i[\bm G ,H] / \hbar \right) \nonumber \\
&&=(2-g)\frac{ mq \hbar}{2E^3}(1 -\frac{1}{2} (\bm \Omega \times \bm x)^2)\left(\bm\sigma-\frac{\bm\sigma\cdot\bm{p}}{E(E+m)} \bm p\right)\cdot(\bm\Omega\times \bm{B})
\label{ndr}
\end{eqnarray}
Details of this cumbersome calculation are given in Appendix \ref{app1}.
The subscript $V$ indicates that the canonical volume form is factored out: $(\frac{1}{2}d\tilde{\omega}_t \wedge \tilde{\omega}_t^2)_{\ssV} \equiv (\frac{1}{2}d\tilde{\omega}_t \wedge \tilde{\omega}_t^2) /{d^3V\wedge dt}.$ 
The other subscript, $M,$ denotes that the Maxwell equations in rotating coordinates, $\partial \bm B/\partial t= -\bm{\nabla}_{\bm x} \times [ \bm E + (\bm\Omega\times \bm x)\times \bm B] ,$  $\bm{\nabla}_{\bm x} \cdot \bm B=0,$ are employed. The former equation was obtained e.g. in Appendix C of \cite{mism2}, under the condition $|\bm \Omega \times \bm x|\ll c.$  

To reduce the Pfaffian and the dynamical equations, (\ref{Pfaf})-(\ref{mspd}), to spin independent  scalars
we can take their trace. Then,   the Liouville equation is satisfied since the trace of (\ref{ndr}) vanishes:
$\Tr [ d \left(\tilde{\omega}_t\right)^3 ] =0.$
Observe that if we choose $g=2,$ Liouville equation is satisfied identically, even before taking the  trace of (\ref{ndr}). Moreover, for most of the calculations involving distribution functions one approximates $\bm \nu\approx \bm p/E.$ In this case. $\bm D\times \bm \nu =0.$ Hence, we conclude that, for Dirac particles the Liouville equation is satisfied:
\begin{equation}
\label{vanm}
{\cal{L}}_{\tilde v} \tilde{\Omega}=0.
\end{equation} 

To discuss particle number and  current densities  within our formalism, one  has to introduce matrix valued distribution functions. Distribution functions which are matrices carrying spin indices show up 
naturally in the framework of  spin dependent Fermi liquids as it was mentioned in Chapter 9 of \cite{landau}. The semiclassical formulation has been elaborated in the spin basis dictated by  the positive energy solutions given in (\ref{ubas}). In this basis the distribution function    will be a matrix with nonvanishing off-diagonal elements which can hardly have a physical interpretation. However, we can  get rid of them by working in the helicity basis rather than  the spin basis of (\ref{ubas}). Helicity operator commutes with the free Dirac Hamiltonian, so that it is a conserved quantity for the free Dirac particles. In the helicity basis, the distribution function can be expressed through  the right-handed and left-handed distribution functions $f_\ssR$ and  $f_\ssL $ as
$$
f=\begin{pmatrix}
f_\ssR & 0\\
0 & f_\ssL
\end{pmatrix}.
$$ 
This separation is also essential  to establish the connection between the massive and massless cases.

The semiclassical helicity matrix  is defined as
$$ {\lambda}^{\alpha \beta}= {u^\alpha}^\dagger \ (\frac{\bm{\Sigma}\cdot \bm{p}}{p}) \ u^\beta = \frac{\bm{\sigma}\cdot \bm{p}}{p}.$$
Employing the spherical coordinates in momentum space we can diagonalize $\lambda$ by the matrix
$$
R=\begin{pmatrix}

\cos(\frac{\theta}{2}) & -\sin(\frac{\theta}{2}) e^{-i \varphi}\\
\sin(\frac{\theta}{2}) e^{i \varphi} & \cos(\frac{\theta}{2})
\end{pmatrix} .
$$
One can easily observe that $R^\dagger \lambda R= \mathrm{diag}\ (1, -1).$

To accomplish  the continuity equation let us deal with  the distribution function  satisfying the  collisionless Boltzmann equation in the helicity basis:
\begin{equation}
(\tilde{\omega}_\ssh)_\ssH\frac{\partial f}{\partial t}+( \dot{\tilde{ \bm x}}\tilde{\omega}_\ssh)_\ssH  \cdot\frac{\partial f}{\partial \bm x}+ (\tilde{\omega}_\ssh \dot{\tilde {\bm p}})_\ssH  \cdot \bm D^\ssH f =0.
\label{boltzmann}
\end{equation}
The subscript $H$ denotes the matrices  written in the helicity basis, like 
$(\tilde{\omega}_\ssh)_\ssH\equiv R^\dagger \tilde{\omega}_\ssh R.$ 
By making use of (\ref{lievolume2}) and (\ref{vanm}), we obtain
$$
 \int \frac{d^3 p}{(2\pi \hbar)^3}\left(\frac{\partial}{\partial t}( (\tilde{\omega}_\ssh)_\ssH f )+\frac{\partial}{\partial \bm{x}}
 \cdot (( \dot{\tilde{ \bm x}}\tilde{\omega}_\ssh )_\ssH f )+\bm D^\ssH \cdot ((\tilde{\omega}_\ssh \dot{\tilde {\bm p}})_\ssH f )\right)     = 0 .
$$
The measure in  phase-space integrals is proportional to the Pfaffian $ \tilde{\omega}_\ssh ,$
so that the probability density is $\rho(x,p,t)= (\tilde{\omega}_\ssh)_\ssH f.$ 
Therefore, the particle number density and the particle current density are given by 
\begin{eqnarray}
n(x,t) &=& \int \frac{d^3p}{(2\pi \hbar)^3} \Tr \left[ (\tilde{\omega}_\ssh)_\ssH f \right] , \label{n,n}  \\
\bm j (x,t) &=& \int \frac{d^3p}{(2\pi \hbar)^3} \Tr \left[( \dot{\tilde{ \bm x}}\tilde{\omega}_\ssh )_\ssH f\right]  .
\label{n,j}
\end{eqnarray}
By setting  $\int \frac{d^3p}{(2\pi \hbar)^3}\bm D \cdot((\tilde{\omega}_\ssh \dot{\tilde {\bm p}})_\ssH f )=0, $  the continuity equation follows,
$$
\frac{\partial }{\partial t}n (x,t)  + \bm {\nabla} \cdot \bm j (x,t)   =0.
$$
Hence,  the particle number  is conserved. Obviously, to reach this conclusion, one can also work in the spin basis. However, the helicity basis is suitable to inspect the vanishing mass limit as well as to take into account chiral imbalance.

\section{Semiclassical Formulation of  Weyl Particles}
\label{hamwe}

The massless limit of the Dirac equation leads to two  equations describing the  Weyl particles. After solving one of these two component Weyl equations, one can compute the related Berry gauge field.
Then,   to establish its semiclassical kinetic theory one proceeds as in the massive case. However, our formulation of the Dirac particles in the helicity basis  directly provides the semiclassical kinetic theory of the Weyl particles either left- or right-handed. In fact by expressing  (\ref{berrycurvature}) in the helicity basis and setting $m=0,$ we readily acquire the Berry curvature 
\begin{equation}
\label{GW}
\bm G_\ssR \equiv \bm b=\hbar\frac{\bm{p}}{2 p^3}.
\end{equation}
We focus on the right-handed Weyl particle.
The symplectic two-form which is the main ingredient of the Hamiltonian approach is deduced from  (\ref{wtf}), as 
\begin{eqnarray}
\omega_t & = & {dp}_i \wedge {dx}_i + \frac{1}{2} \epsilon_{ijk} (q B_k + 2 {\cal E}_0 \Omega_k) {dx}_i \wedge {dx}_j + \epsilon_{ijk}   \Omega_k x_j \nu_{\sif m} {dx}_i \wedge {dp}_m  \nonumber \\ 
&&- \frac{1}{2} \epsilon_{ijk} b_{k} {dp}_i \wedge {dp}_j + \frac{1}{2}  \nu_{\sif i} (\bm \Omega \times \bm x)^2 {dp}_i \wedge dt \nonumber \\ 
&&-  \nu_{\sif i} {dp}_i \wedge dt 
+[q\bm E+ (\bm\Omega\times \bm x)\times (q\bm B +{\cal E}_0 \bm\Omega)]_i{dx}_i \wedge dt .
\label{masslesstwoform}
\end{eqnarray}
${\bm \nu}_0$ is delivered by  (\ref{numas}) after expressing it in the helicity basis and setting $m=0:$ 
\begin{eqnarray}
\label{nu0}
{\bm \nu}_0&=&\frac{\bm p}{p} + g\hbar \frac{\bm p}{2 p^3} ( \bm{\Omega}\cdot\bm{p} ) - g\hbar \frac{\bm \Omega}{2 p} + \hbar q \frac{\bm p}{p^4} (\bm{B} \cdot \bm{p}) - \hbar q \frac{\bm B}{2 p^2}.
\end{eqnarray}
On the other hand, ${\bm \nu}_0$ could also be defined by
the semiclassical Weyl  Hamiltonian  $H_0,$ which is  established from (\ref{Hambo}) in the helicity basis as
\begin{equation}
\label{hammw}
H_0 \equiv {\cal E}_0= p - g p^2 (\bm{b} \cdot \bm{\Omega}) -  q p (\bm{B} \cdot \bm{b}) .
\end{equation}
Indeed, ${\bm \nu}_0 = \partial H_0/ \partial \bm p $   produces (\ref{nu0}).

To derive the equations of motion let us introduce the vector field
\begin{equation}
\label{vfw}
v= \frac{\partial}{\partial t}+\dot{ \bm{x}}\frac{\partial}{\partial \bm{x}}+\dot{\bm {p}}\frac{\partial}{\partial \bm{p}}.
\end{equation}
The interior product of the vector field (\ref{vfw}), with the 
symplectic two-form (\ref{masslesstwoform}),
\begin{equation}
\label{eqmw}
i_{v} \omega_t = 0,
\end{equation}
yields the following coupled equations of motion
\begin{eqnarray}
\dot{\bm x} +  (\bm \Omega \times \bm x) \cdot \dot{\bm x}\ \bm \nu_0 &=& \bm \nu_0 (1 -\frac{1}{2} (\bm \Omega \times \bm x)^2) +  \dot{\bm p} \times \bm b  ,\label{c1} \\
\dot{\bm p}  + ( \dot{\bm p} \cdot \bm \nu)\ \bm \Omega \times \bm x &=& q\bm E + (\bm\Omega\times \bm x)\times (q\bm B +{\cal E}_0 \bm\Omega) \ +  \dot{\bm x} \times  (q\bm B +2 {\cal E}_0 \bm \Omega ).\label{c2}
\end{eqnarray}
These could also be obtained  from the massive equations of motion (\ref{xeq}),(\ref{peq}). It is worth noting that the second term in (\ref{c2}) is the centrifugal-like force for Weyl particles.

\section{Semiclassical Transport of Weyl Particles}
\label{trawe}

To  discuss  particle number conservation for the Weyl particles, we need to attain the related Liouville equation
which can be accomplished by calculating  the Lie derivative  of the volume form 
\begin{equation}
\label{omm}
\Omega =\frac{1}{3!} {\omega}_t^3 \wedge dt= \frac{1}{3!} {\omega}^3 \wedge dt,
\end{equation}
where $\omega\equiv\omega_t|_{dt=0}$ is the symplectic form in the six-dimensional ordinary phase-space. 
As we have already seen in the massive case, by expressing the Lie derivative of volume form in two different ways and  comparing them,  one can solve  the coupled equations of motion (\ref{c1})-(\ref{c2}) for the velocities $(\dot{\bm x}, \dot{\bm p}),$ in terms of the phase-space variables $(\bm x, \bm p).$ 
By making use of  (\ref{eqmw}), the Lie derivative  can be calculated as
\begin{eqnarray}
{\cal{L}}_{v} \Omega &=&  (i_{v}  d + d i_{v} )(\frac{1}{3!} {{\omega}_t}^3 \wedge dt)\nonumber\\
&=& \frac{1}{3!} d{\omega_t}^3.
\label{rli0}
\end{eqnarray}
On the other hand   by means of $\sqrt{\omega}, $ which is the Pfaffian of the symplectic matrix (\ref{syma}), where  $\bm G$ is  given by (\ref{GW}), the volume form (\ref{omm}) can be expressed as
\begin{equation}
\label{WO}
\Omega =\sqrt{\omega} dV \wedge dt.
\end{equation}
Then the Lie derivative of (\ref{WO}) yields
\begin{eqnarray}
{\cal{L}}_{v} \Omega &=&  \left( \frac{\partial \sqrt{\omega}}{\partial t} + \frac{\partial (\sqrt{\omega} \dot{{\bm x}})}{\partial \bm{x}} + \frac{\partial (\sqrt{\omega} \dot{{\bm p}})}{\partial \bm{p}}\right) dV \wedge dt.
\label{lievol0}
\end{eqnarray}
Now, by calculating ${\omega_t}^3$ and  comparing (\ref{rli0}) with (\ref{lievol0}),  Pfaffian and
the velocities of  phase-space variables  are revealed to be
\begin{eqnarray}
\sqrt{\omega} &=&1+  \ \bm{b} \cdot (q\bm{B} + 2 p \bm{\Omega} ) - \bm \nu_{0} \cdot (\bm x \times \bm \Omega) - (  \hat{\bm p}  \cdot \bm b)(q \bm B \cdot (\bm x \times \bm \Omega)),
\label{smpf} \\
\sqrt{\omega}  \dot{{\bm x}}&=&{\bm \nu}_0 (1 -\frac{1}{2} (\bm \Omega \times \bm x)^2) +  \ \bm e \times \bm b \nonumber\\
&&+  (\hat{\bm p} \cdot \bm b) (q \bm B+ 2p\bm \Omega)(1  -\frac{1}{2} (\bm \Omega \times \bm x)^2)+( \hat{\bm p} \cdot \bm b) [ (\bm x \times \bm \Omega) \times \bm e ]    ,\label{smlx}\\
\sqrt{\omega}  \dot{{\bm p}} &=& \bm e+ {\bm \nu}_0 \times (q\bm{B} + 2 p \bm{\Omega}) (1  -\frac{1}{2} (\bm \Omega \times \bm x)^2) \nonumber\\
&& + \ \bm b (\bm e \cdot (q \bm{B} + 2p \bm{\Omega})) - [ (\bm x \times \bm \Omega) \times \bm e ] \times {\bm \nu}_0 .  \label{smlp}
\end{eqnarray}
$\bm e$ is given as in (\ref{efel}) for $m=0:$  $\bm e =  q\bm E + (\bm\Omega\times \bm x)\times (q\bm B +{\cal E}_0 \bm\Omega).$ 
These solutions can also be detected from the massive ones (\ref{Pfaf})-(\ref{mspd}) by setting $m=0$ and retaining the right-handed part in the helicity basis.

Some other solutions for velocities of phase-space variables  and the Pfaffian were proposed
in \cite{cpww}. They  start from a Lorentz invariant quantum Boltzmann equation and reduce it to three-dimensions.
If we ignore  centrifugal terms and the last term,  (\ref{smlx})  coincides with $\sqrt{\omega}  \dot{{\bm x}}, $ presented in \cite{cpww}. However, the others differ. In our formalism  the symmetry between ${2p}\bm \Omega$ and $\bm B$ is respected up to centrifugal terms; however, this  is not the case in   \cite{cpww}: When we switch off the electric field $\bm E,$ and ignore  centrifugal terms,  (\ref{smlp}) leads to 
$\sqrt{\omega}  \dot{{\bm p}} = \hat{\bm p} \times (q\bm{B} + 2 p \bm{\Omega}) . $ This reflects the fact that there is a Coriolis   forcelike term for Weyl particles which vanishes for $2p \bm \Omega =-q\bm B,$ similar to the Dirac particles. But $\sqrt{\omega}  \dot{{\bm p}} ,$ acquired in \cite{cpww} yields only the magnetic force for vanishing electric field. For $\bm \Omega =0,$  either our solutions (\ref{smpf})-(\ref{smlp}) or the ones given  in \cite{cpww} reproduce the solutions obtained in \cite{sy}.

The Lie derivative of the volume form can be calculated either by inserting the solutions (\ref{smpf})-(\ref{smlp}) into (\ref{lievol0}) or in terms of the symplectic  form    (\ref{masslesstwoform}) as
\begin{eqnarray}
&(\frac{1}{2}dw_t \wedge {w_t}^2 )_{\ssV\ssM} &
= 2\pi \hbar q^2\delta(\bm p)\bm{{ E}}\cdot \bm{B}+4\pi \hbar q\delta(\bm p)\bm{{ E}}\cdot (p \bm{\Omega}) -2\pi \hbar q\delta(\bm p)\bm{B}\cdot((\bm{\Omega}\times\bm{x})\times p \bm{\Omega}) \nonumber \\
&& =  2\pi \hbar q^2\delta(\bm p)\bm{{ E}}\cdot \bm{B}.\label{vma}
\end{eqnarray}
In Appendix \ref{app1}, we presented the details of this calculation.
The subscript $VM$ denotes that the canonical volume form $dV\wedge dt$ is factored out and Maxwell equations $\nabla_{\bm x} \cdot \bm B=0$ and $ \nabla_{\bm x} \times \bm e =- q\partial \bm B/\partial t,$ have been employed.
To exhibit the role of  monopole located  at the origin we presented the second and third terms which are actually vanishing. We conclude that the  Liouville equation is anomalous.

To inspect the kinetic theory let us introduce the distribution function for the right-handed fermions, $f_\ssR,$ satisfying the  Boltzmann equation  without collisions:
\begin{equation}
\frac {df_\ssR}{dt}=\frac{\partial f_\ssR}{\partial t}+\frac{\partial f_\ssR}{\partial \bm{x}}\cdot \dot{\bm{x}}+\frac{\partial f_\ssR}{\partial \bm{p}}\cdot \dot{\bm{p}}=0. \nonumber 
\end{equation}
Therefore we get
\begin{eqnarray}
\int \frac{d^3 p}{(2\pi \hbar)^3}\left(\frac{\partial}{\partial t}( \sqrt{\omega} f_\ssR )+\frac{\partial}{\partial \bm{x}}
\cdot ( \sqrt{\omega}\dot{\bm x} f_\ssR )+\frac{\partial}{\partial \bm p} \cdot (\sqrt{\omega} \dot{\bm p} f_\ssR )\right)     = \int \frac{d^3 p}{(2\pi \hbar)^3} ( \frac{1}{2}d\omega_t \wedge {\omega_t}^2)_{\ssV\ssM} f_\ssR. \nonumber              
\label{coneq0}
\end{eqnarray}

Measure of the phase-space integrals differs from the canonical value up to the Pfaffian $\sqrt{\omega}.$
Hence,   particle number and current densities are defined by
\begin{eqnarray}
	n_\ssR (x,t) &=& \int \frac{d^3p}{(2\pi \hbar)^3} \sqrt{\omega} f_\ssR  , \label{WNO} \\
	\bm j_\ssR (x,t) &=& \int \frac{d^3p}{(2\pi \hbar)^3} \sqrt{\omega}\dot{ \bm x}f_\ssR  .\label{WJO}
\end{eqnarray}
The 4-divergence of the particle 4-current $(n_\ssR,\bm j_\ssR )$ can be written as
\begin{equation} 
\label{ceqD0}
\frac{\partial n_\ssR}{\partial t} + \bm {\nabla \cdot j_\ssR} = \int \frac{d^3 p}{(2\pi \hbar)^3}  (\frac{1}{2}d\omega_t \wedge {\omega_t}^2)_{\ssV\ssM} f_\ssR.    
\end{equation}
We then conclude that Weyl particles satisfy the following  continuity equation with  source, 
\begin{equation}
\label{a1}
\frac{\partial n_\ssR (\bm x,t)}{\partial t}+\bm{\nabla} \cdot \bm{j}_\ssR (\bm x,t)=\frac{ q^2}{4 \pi^2 \hbar^2}  \bm E \cdot \bm B.
\end{equation}
Following the same procedure for the left-handed fermions we accomplish
\begin{equation}
\label{a2}
\frac{\partial n_\ssL (\bm x,t)}{\partial t}+\bm{\nabla} \cdot \bm{j}_\ssL (\bm x,t)=-\frac{ q^2}{4 \pi^2 \hbar^2}  \bm E \cdot \bm B . 
\end{equation}
Observe that (\ref{n,n})-(\ref{n,j}) yield (\ref{WNO})-(\ref{WJO})   in the vanishing mass limit. 
The continuity equations (\ref{a1}) and (\ref{a2}) are consistent with the results of generic chiral hydrodynamics accomplished in \cite{ss}.

\section{The Anomalous Chiral Effects}
\label{ace}
The anomalous chiral transport effects and the  related experimental results were recently  reviewed in   \cite{khetal-arx}. In the light of  recent experiments it is asserted that
most probably the anomalous chiral effects show up in  heavy ion collisions. 
Let us analyze how  our semiclassical approach produces  the  anomalous chiral effects due to external electromagnetic fields and global rotation.

We can readily consider  the $m=0$ limit of the current generated by Dirac particles  in the  helicity basis  (\ref{n,j}), which can be written as
$$
\bm j= \begin{pmatrix} 
\bm j_\ssR & 0 \\
0 & \bm j_\ssL
\end{pmatrix} .
$$
One  can observe that $\bm j_\ssR,$ coincides with the current (\ref{WJO}), which is given by 
\begin{eqnarray}
\label{jr}
\bm j_\ssR  &=& \int \frac{d^3p}{(2\pi \hbar)^3} \Big({\bm \nu}_0 (1  -\frac{1}{2} (\bm \Omega \times \bm x)^2) +   \bm e \times \bm b \nonumber\\
&&+  ({\bm \nu}_0 \cdot \bm b) (q \bm B+ 2p \bm \Omega)(1  -\frac{1}{2} (\bm \Omega \times \bm x)^2)+ ( \bm \nu_{0} \cdot \bm b) [ (\bm x \times \bm \Omega) \times \bm e ]    \Big) f_\ssR . 
\end{eqnarray}
Similarly we can derive the current
for the particles of positive helicity as
\begin{eqnarray}
\label{jl}
\bm j_\ssL  &=& \int \frac{d^3p}{(2\pi \hbar)^3} \Big({\bm \nu}_0^\ssL (1 -\frac{1}{2} (\bm \Omega \times \bm x)^2) - \bm e \times \bm b \nonumber\\
&&-  ({\bm \nu}_0^\ssL \cdot \bm b) (q \bm B+ 2p \bm \Omega)(1  -\frac{1}{2} (\bm \Omega \times \bm x)^2)- ( \bm \nu_{0}^\ssL \cdot \bm b) [ (\bm x \times \bm \Omega) \times \bm e ]    \Big) f_\ssR . 
\end{eqnarray}

For the left-handed fermions we introduced
\begin{eqnarray*}
	{\bm \nu}^\ssL_0&=&\frac{\bm p}{p} - g\hbar \frac{\bm p}{2 p^3} ( \bm{\Omega}\cdot\bm{p} ) +g \hbar \frac{\bm \Omega}{2 p} - \hbar q \frac{\bm p}{p^4} (\bm{B} \cdot \bm{p}) + \hbar q \frac{\bm B}{2 p^2}.
\end{eqnarray*}

In terms of the right- and left-handed particle number current densities $\bm j_\ssR$ and  $\bm j_\ssL,$
one defines the vector and axial currents:
$$
\bm j_\ssV =\bm j_\ssR +\bm j_\ssL,\ \  \bm j_\ssA =\bm j_\ssR -\bm j_\ssL .
$$
Let us
deal with the right- and left-handed fermions obeying Fermi-Dirac distribution whose  respective  chemical potentials are denoted  by  $\mu_\ssR,$ and $\mu_\ssL :$  $f_{\ssR (\ssL)}=f_{\ssF\ssD}(E,\mu_{\ssR (\ssL)}).$
We ignore  quantum corrections to the energy (\ref{hammw}), so that
we  can perform the integrals in (\ref{jr}) and (\ref{jl}) by setting $\bm \nu_0=\bm \nu_0^\ssL =\bm{\hat{p}}.$
In the integrals we set the surface terms equal to zero and make use of the approximation \cite{abr} valid for a well-behaved function $F(E):$
$$
\int F(E)\frac{\partial f_{\ssF\ssD}(E, \mu_{\ssR (\ssL)}) }{\partial E}dE \approx -F (\mu_{\ssR (\ssL)} ) -
\frac{1}{6} \pi^2 T^2 \frac{\partial ^2F(E)}{\partial E^2}|_{E=\mu_{\ssR (\ssL)}}.
$$
The chiral magnetic effect  and the  chiral separation effect
are  generated by the
terms which are proportional to the magnetic field $\bm B,$ respectively, in the vector and axial currents.  They are calculated to be 
\begin{eqnarray*}
	\bm j_\ssV^\cme &=&\frac{ q }{2\pi^2 \hbar^2} \mu_5 \bm B, \\
	\bm j_\ssA^{\ssC \ssS \ssE} &=&\frac{ q }{2\pi^2 \hbar^2} \mu \bm B.
\end{eqnarray*}
We  introduced  the  total chemical potential $\mu =\frac{1}{2}(\mu_\ssR+\mu_\ssL )$ and the chiral chemical potential  $\mu_5 =\frac{1}{2}(\mu_\ssR-\mu_\ssL) .$
On the other hand, the
terms which are proportional to  the angular velocity $\bm \Omega,$
in the vector and axial currents, respectively, generate
the chiral vortical effect and the local (spin) polarization effect. They are obtained as
\begin{eqnarray*}
	\bm j_\ssV^\cve &=&\frac{\mu\mu_5}{ \pi^2 \hbar^2}  \bm \Omega \\
	\bm j_\ssA^{\ssL \ssP \ssE} &=& \left\{\frac{1 }{2\pi^2\hbar^2} (\mu^2 +\mu_5^2)+\frac{T^2}{6\hbar^2}\right\} \bm \Omega.
\end{eqnarray*}
These are in accord with the results obtained within other approaches (see \cite{khetal-arx} and the references therein). Temperature dependence has been  obtained e.g. in \cite{32} which was then shown
to be related to the mixed gauge-gravity anomaly \cite{33}. 

\section{Discussions}

The semiclassical formalism which we presented provides an intuitive  understanding of the transport phenomena of the Dirac and Weyl particles in the presence of the external electromagnetic fields as well as rotation of the coordinate frame. It delivers the anomalous chiral effects straightforwardly. 
We considered noninteracting particles, though the kinetic theory is powerful in studying transport equations in the presence of interactions \cite{cint}, \cite{vint}. 
The results obtained here should be considered  as the first step in that direction.
Our formulation of the Dirac particles has the advantage of exhibiting spin degrees of freedom explicitly, thus it is adequate to deal with spin dependent interactions. 
Therefore the exposed straightforward connection between the massive and massless cases
can give clues about the systematic study of chirality imbalance.

It is known that rotations of coordinates generate spin currents due to spin-rotation coupling \cite{mism2,pap,imm}.
 For  condensed matter systems 
spin currents  
are mostly studied  by considering the third component of spin, although generally spin is not a conserved quantity. However the helicity basis is suitable for  analyzing  spin currents and  calculating the related spin Hall conductivities \cite{om-elif}. 
Therefore  the semiclassical kinetic theory of the Dirac particles 
developed here will yield a better understanding of spin currents generated by  rotations of coordinate frames.

\begin{acknowledgments}
		This work is supported by the Scientific and Technological Research Council of Turkey (T\"{U}B\.{I}TAK) Grant No. 115F108.
\end{acknowledgments}

\appendix

\renewcommand{\theequation}{\thesection.\arabic{equation}}
\setcounter{equation}{0}

\section{The Lie Derivatives of  Volume Forms for Dirac and Weyl Particles }
\label{app1}
The  Lie derivative of  volume form suitable to Dirac particles can   be computed directly as
\begin{eqnarray}
{\cal{L}}_{\tilde v} \tilde{\Omega} 
&=& \frac{1}{2} d \tilde{\omega}_t \wedge \tilde{\omega}_t \wedge \tilde{\omega}_t .
\label{liouvilleA}
\end{eqnarray}
When the exterior derivative of the symplectic two-form (\ref{wtf}) is taken there are some terms which cancel each other: $(i)$ the term that includes the covariant derivative of the effective electric field, $ D_j e_i \ dp_j\wedge dx_i \wedge dt$ 
and $\nu_i((\bm \Omega \times \bm x)\times \bm x)_j dx_j \wedge dp_i \wedge dt$, which is the spatial derivative of \mbox{$\frac{\nu_i}{2}(\bm \Omega \times \bm x)^2 dp_i \wedge dt,$} term in $\tilde{ \omega}_t,$ cancel each other, and $(ii)$ $D_l{\cal E}\epsilon_{ijk}\Omega_k dp_l\wedge dx_i \wedge dx_j$ cancels out  $\epsilon_{ijk}\Omega_j\nu_j dx_k \wedge dp_j \wedge dx_i$, which arises from  the spatial derivative of $\nu_m (\bm \Omega \times \bm x)_i dp_m \wedge dx_i$ term in (\ref{wtf}). Thus, the only nonvanishing terms are 
\begin{eqnarray}
	d\tilde{\omega}_t&=&-\frac{1}{2}\epsilon_{ijk}D_lG_k \  {dp}_l \wedge dp_i \wedge dp_j- \left(1-\frac{1}{2}(\bm \Omega \times \bm x)^2 \right) D_i \nu_j dp_k \wedge dp_i \wedge dt \nonumber\\ 
	&&+\frac{q}{2}\epsilon_{ijk}\frac{\partial B_k}{\partial x_l}  \  dx_l \wedge dx_i  \wedge dx_j+D_i \nu_j \epsilon_{klm}\Omega_l x_m dp_i \wedge dp_j \wedge dx_k \nonumber \\
	&&+\frac{\partial e_i}{\partial x_k} dx_k \wedge dx_i \wedge dt+\frac{q}{2}\epsilon_{ijk}\frac{\partial B_k}{\partial t}dx_i \wedge dx_j \wedge dt. \label{dwmass} \nonumber
	\end{eqnarray}   
In the wedge product of $d\tilde{\omega}_t$ with ${\omega_t}\wedge \omega_t,$ most of the terms vanish due to antisymmetry of the wedge product.
Thus the nonvanishing terms of (\ref{liouvilleA}) are  as follows, 
\begin{eqnarray}
\frac{1}{2}d\tilde{\omega}_t \wedge \tilde{\omega_t}^2&=&\Bigg\{(\bm D\cdot\bm G)(\bm B+2{\cal E} \bm\Omega)\cdot\bm e+\left[1-\frac{1}{2}(\bm \Omega \times \bm x)^2\right] (\bm{D}\times\bm{\nu})\cdot (\bm B+2{\cal E} \bm\Omega)\nonumber\\
&&+(\bm \nabla \cdot \bm B) (\bm \nu \cdot \bm G)+(q\frac{\partial \bm B}{\partial t}+\bm \nabla \times \bm e)\cdot
 \left[ \bm G+(\bm{\nu}\cdot\bm G) (\bm \Omega \times \bm x) \right] 
\Bigg\}d^3V\wedge dt. 
\label{dwwsq}
\end{eqnarray}
The first term vanishes due to $\bm{ D} \cdot \bm G =0.$ Using the Maxwell equations in rotating coordinates, the last line also vanishes. Then, we conclude that (\ref{dwwsq}) leads to (\ref{ndr}):
\begin{eqnarray*}
\frac{1}{2}d\tilde{\omega}_t \wedge \tilde{\omega}_t^2 = \Bigg(
(1 -\frac{1}{2} (\bm \Omega \times \bm x)^2)(q\bm B+2E\bm \Omega) \cdot
(\bm{D}\times \bm{\nu})\Bigg) d^3V\wedge dt. 
\end{eqnarray*}
 
Similar calculations can be done for the massless case by using the symplectic two-form (\ref{masslesstwoform}). 
While taking the exterior derivative of (\ref{masslesstwoform}), cancellations mentioned before (\ref{dwmass})  occur in the massless case, too. Thus the  exterior derivative of (\ref{masslesstwoform}) leads to
\begin{eqnarray}
d\tilde{\omega}_t&=&- \frac{1}{2} \frac{\partial b_k}{\partial p_l} \epsilon_{ijk}\  {dp}_l \wedge dp_i \wedge dp_j-\frac{\partial \nu_{\sif i}}{\partial p_k}\left(1-\frac{1}{2}(\bm \Omega \times \bm x)^2\right) dp_k \wedge dp_i \wedge dt  \nonumber\\ 
&&+\frac{q}{2}\epsilon_{ijk}\frac{\partial B_k}{\partial x_n}  \  dx_n \wedge dx_i  \wedge dx_j + \frac{\partial \nu_{\sif i}}{\partial p_k}\epsilon_{lmn}\Omega_m x_n dp_k \wedge dp_i \wedge dx_l \nonumber \\
&&+\frac{\partial e_i}{\partial x_k} dx_k \wedge dx_i \wedge dt+\frac{q}{2}\epsilon_{ijk}\frac{\partial B_k}{\partial t}dx_i \wedge dx_j \wedge dt.
\label{dwmassless}
\end{eqnarray}

Both of the $ \partial \nu_{\sif i}/\partial p_k$ terms in (\ref{dwmassless}), multiplied with the terms containing  the four-form $dx^3 dp$ in ${\omega_t}\wedge \omega_t,$ vanish due to antisymmetry of wedge product.
Hence, the Lie derivative of the volume form for Weyl particles is obtained as
\begin{eqnarray}
{\cal{L}}_{ v} \Omega=&& \bm e \cdot (q \bm B + 2 p \bm \Omega) (\bm \nabla \cdot \bm b)\nonumber \\
&&+(\bm \nabla \cdot \bm B) (\bm \nu_{0} \cdot \bm b)+  \left(\bm \nabla \times \bm e + \frac{\partial \bm B}{\partial t} \right) \cdot \bm \{ \bm b -(\bm x \times \bm \Omega) (\bm \nu_{0} \cdot \bm b)\bm \} 
d^3V\wedge dt.
\label{anomalymassless}
\end{eqnarray}
 Obviously, the last two terms  in (\ref{anomalymassless}) vanish when  the Maxwell equations in rotating coordinates are satisfied. Finally, we obtain
$$
\frac{1}{2}d\omega_t \wedge {\omega_t}^2 =  2\pi \hbar q^2\delta(\bm p)\bm{{ E}}\cdot \bm{B}d^3V\wedge dt,
$$
which is the result given in  (\ref{vma}).

\section{Comparison With the Pauli-Schroedinger Hamiltonian Approach} 
\label{app2}

We would like to compare the force  which we acquired in terms of the wave packet composed of the positive energy solutions of the Dirac equation with the one presented in \cite{mism2}, where the Pauli-Schr\"{o}dinger Hamiltonian was employed. By ignoring the terms at  $\hbar^2$ order, they accomplished the following force,
\begin{eqnarray}
\label{mi}
&&\bm F =q\bm{E}'+q\bm v \times \bm B +2m\bm v \times \bm\Omega-m\bm\Omega\times(\bm\Omega\times \bm x)
-\frac{q\hbar}{4m^2}\Big[(\bm{\sigma}\times\bm{E}')\times(q\bm{B}+m\bm{\Omega})  \nonumber \\
&&-((q\bm{B}
+m\bm{\Omega})\times\bm{\sigma})\times\bm{E}'\Big] +\frac{q\hbar}{4m}\Big[\bm{\sigma}\cdot(\bm{\Omega}\times\bm{v})\bm{B}
-2(\bm{\Omega}\cdot\bm{B})\bm{\sigma}\times\bm{v}-(\bm{B}\cdot\bm{v})\bm{\sigma}\times\bm{\Omega}  \nonumber \\
&&+\bm{\Omega}\cdot(\bm{x}\times\bm{B})\bm{\sigma}\times\bm{\Omega}
-(\bm{B}\cdot\bm{\Omega})\bm{\sigma}\times(\bm{\Omega}\times\bm{x})\Big],
\end{eqnarray}
where $\bm{E}'=\bm{E}+(\bm\Omega\times \bm x)\times \bm B$ is the effective electric field 
in a rotating frame when  angular speed is much smaller than the speed of light.

Within our approach the force acting on the Dirac particle  in  nonrelativistic limit can be established from 
(\ref{mspd}) ignoring the last term which gives rise to higher-order terms in momentum.  We also ignore  the term proportional to  $(\bm{\Omega}\times\bm{x})^2,$ which is irrelevant for the comparison with (\ref{mi}). We invert the Pfaffian at the first order in $\hbar$ as $\tilde{\omega}^{-1}\approx 1-\bm{G}\cdot (q \bm{B} + 2E \bm{\Omega}),$ neglecting the last two terms in (\ref{Pfaf}) since they are higher order terms in momentum. Thus, we acquire the force  as 
$$
\dot{\tilde {\bm p}} = \bm e+\bm{\nu} \times (q\bm{B} + 2E \bm{\Omega}) - (q \bm{B} + 2E \bm{\Omega})\times( \bm G \times\bm e ) +\bm G \cdot (q\bm{B} + 2E \bm{\Omega})[\bm{\nu}\times (q\bm{B} + 2E \bm{\Omega})].
$$
In the limit $E\rightarrow m$, and setting the velocity $\bm v =\bm p / m,$ it leads to 
\begin{eqnarray*}
	&&\dot{\tilde {\bm p}}= q\bm{E}'+q\bm v \times \bm B +2m\bm v \times \bm\Omega  -m\bm\Omega\times(\bm\Omega\times \bm x) -\frac{q\hbar}{2m^2}(\bm{\sigma}\times\bm{E}')\times(q\bm{B}+2m\bm{\Omega})\\
	&&-\frac{q\hbar}{2m}\Big[2(\bm{\sigma}\cdot(\bm{\Omega}\times\bm{v}))\bm{B}+g(\bm{\Omega}\cdot\bm{B})\bm{\sigma}\times\bm{v}
	-g(\bm{B}\cdot\bm{v})\bm{\sigma}\times\bm{\Omega}-(2-g)(\bm{\sigma}\cdot(\bm{B}\times\bm{v}))\bm{\Omega}\\
	&&+(2-g)\bm{v}(\bm{\sigma}\cdot(\bm{B}\times\bm{\Omega}))-\bm{\Omega}\cdot(\bm{x}\times\bm{B})\bm{\sigma}\times\bm{\Omega} +(\bm{B}\cdot\bm{\Omega})\bm{\sigma}\times(\bm{\Omega}\times\bm{x})\Big] \\
	&&+\frac{q^2\hbar}{2m^2}(\bm{\sigma}\cdot\bm{B}) \bm{v}\times\bm{B}+\hbar(g-2)(\bm{\sigma}\cdot\bm{\Omega}) \bm{v}\times\bm{\Omega} +\frac{q\hbar}{2m}\Big[(\bm{\Omega}\times(\bm{\Omega}\times\bm{x})\times(\bm{\sigma}\times\bm{B})\Big]\\
	&&+\hbar\bm{\Omega}\times\Big[\bm{\sigma}\times((\bm\Omega\times \bm x)\times \bm{\Omega})\Big]+\frac{\hbar}{4m^2}\bm{\sigma}\cdot\bm{v}(q\bm{v}\cdot\bm{B}+m(g-2)\bm{v}\cdot\bm{\Omega})\bm{v}\times(q\bm{B}+2m\bm{\Omega})\\
	&&+\frac{\hbar}{4m^2}(q\bm{B}+2m\bm{\Omega})\times\Big[(\bm{\sigma}\cdot\bm{v})\bm{v}\times(q\bm{E}'+m(\bm\Omega\times \bm x)\times \bm{\Omega})\Big].
\end{eqnarray*}
Neglecting higher-order terms in the velocity to compare with (\ref{mi}), we obtain
\begin{eqnarray}
	&&\dot{\tilde {\bm p}}= q\bm{E}'+q\bm v \times \bm B+2m\bm v \times \bm\Omega-m\bm\Omega\times(\bm\Omega\times \bm x) -\frac{q\hbar}{2m^2}(\bm{\sigma}\times\bm{E}')\times(q\bm{B}+2m\bm{\Omega}) \nonumber\\
	&&-\frac{q\hbar}{2m}\Big[2(\bm{\sigma}\cdot(\bm{\Omega}\times\bm{v}))\bm{B}+g(\bm{\Omega}\cdot\bm{B})\bm{\sigma}\times\bm{v}
	-g(\bm{B}\cdot\bm{v})\bm{\sigma}\times\bm{\Omega}-(2-g)(\bm{\sigma}\cdot(\bm{B}\times\bm{v}))\bm{\Omega}\nonumber\\
	&&+(2-g)\bm{v}(\bm{\sigma}\cdot(\bm{B}\times\bm{\Omega}))-\bm{\Omega}\cdot(\bm{x}\times\bm{B})\bm{\sigma}\times\bm{\Omega} +(\bm{B}\cdot\bm{\Omega})\bm{\sigma}\times(\bm{\Omega}\times\bm{x})\Big]
	\label{fus} \\
	&&+\frac{q^2\hbar}{2m^2}(\bm{\sigma}\cdot\bm{B}) \bm{v}\times\bm{B}+\hbar(g-2)(\bm{\sigma}\cdot\bm{\Omega}) \bm{v}\times\bm{\Omega} +\frac{q\hbar}{2m}\Big[(\bm{\Omega}\times(\bm{\Omega}\times\bm{x})\times(\bm{\sigma}\times\bm{B})\Big]\nonumber\\
	&&+\hbar\bm{\Omega}\times\Big[\bm{\sigma}\times((\bm\Omega\times \bm x)\times \bm{\Omega})\Big] \nonumber
\end{eqnarray}
The terms at the order $\hbar^0$  coincide. However, the terms dependent on the centrifugal force dependent terms differ.
Although neither $g=1$ nor $g=2$ yield an exact match between (\ref{mi}) and (\ref{fus}), choosing $g=2,$ yields a better match.

In \cite{mism2}, the effects of impurity scatterings were discussed by considering the Dirac particles moving in the plane perpendicular to  $\bm{\Omega},$ which is in the same direction with $\bm{B},$
by neglecting higher-order terms in $\bm{\Omega}$ and taking $q\bm{B}>>m\bm{\Omega}$. Under these conditions  our formulation for $\bm E=0$ leads to 
\begin{eqnarray*}
	&&\dot{\tilde {\bm p}}= q(\bm{B}\cdot\bm{\Omega})\bm{x}-\frac{q^2\hbar}{2m^2}(\bm{\sigma}\cdot\bm{B})(\bm{B}\cdot\bm{\Omega})\bm{x}+q\bm{v}\times\bm{B},
\end{eqnarray*}
which has the same form with the one obtained from (\ref{mi}), except
the second term is smaller by a factor of two.  This discrepancy between   the semiclassical formulations  established by making use of the Dirac and Pauli wave packets, has already  been mentioned in \cite{chni}.

\newcommand{\PRL}{Phys. Rev. Lett. }
\newcommand{\PRB}{Phys. Rev. B }
\newcommand{\PRD}{Phys. Rev. D }

\end{document}